\documentclass[pre,twocolumn]{revtex4-2}
\usepackage{graphicx}
\usepackage{amsopn}
\usepackage{amsfonts}
\usepackage{amsmath}
\usepackage{latexsym}
\usepackage{bm}
\usepackage{amssymb}
\usepackage{color}

\begin{document}

\title{Correlations of the phase gradients of the light wave \\
propagating in a turbulent medium in the regime of strong scintillations}

\author{V.A.Bogachev$^{1}$,
I.V. Kolokolov$^{2}$, V.V. Lebedev$^{2}$, A.V.Nemtseva$^{1,3}$,
and F.A.Starikov$^{1,4}$}

\affiliation{
$^1$Russian Federal Nuclear Center -- Russian Research Institute of Experimental Physics \\
37 Mira ave, Sarov, Nizhny Novgorod region, 607188, Russia;  \\
$^2$ Landau Institute for Theoretical Physics, RAS, \\
142432, Chernogolovka, Semenova 1A, Moscow region, Russia; \\
$^3$Lomonosov State University, Faculty of Physics, \\
Sarov Branch of Lomonosov State University, \\
2 Parkovaya St., Sarov, Nizhny Novgorod Reg., 607328 Russia; \\
$^4$National Research Nuclear University MEPhI -- Sarov Physical and Technical Institute, \\
6 Dukhova ave, Sarov, Nizhny Novgorod region, 607186, Russia. }

\date{\today}

\begin{abstract}

We investigate analytically and numerically correlation functions of the phase of light waves that propagate through turbulent media. We examine the case of strong scintillations that occur at large values of the Rytov dispersion, $\sigma^2_R$. Then, it is possible to relate the pair correlation function of phase gradients to the known pair correlation function of the envelope dependent on the distance $r$ assuming Gaussianity of the envelope of the beam. Our direct numerical simulations show that the profile of the pair correlation function for phase gradients gradually approaches the theoretical expression as the value of $\sigma_R^2$ increases, if $r<r_0$ where $r_0$ is the Fried length. For larger $r$ the behavior of the computed correlation function is quite different because of destroying the Gaussianity.

\end{abstract}


\maketitle

\section{Introduction}
\label{sec:intoduction}

This paper focuses on a comparison between analytical studies and numerical simulations of light beam propagation through a turbulent medium, such as the Earth's atmosphere. The turbulence causes fluctuations in the refractive index of the medium, which leads to distortions in the light beam. Observations and theoretical arguments have shown that the character of these distortions depends on the path taken by the beam. For relatively short distances, the distortions are small, and this is known as the weak scintillation regime. However, for longer distances, the light beam breaks up into individual speckles, which is known as the strong scintillation regime. It is the subject of our work.

The approach used to examine peculiarities of light propagation in a turbulent medium is traced back to the classical works of Kolmogorov and Obukhov \cite{Kolm411,Kolm412,Obukhov41}, based on the concept of energy cascade. In the works, scaling laws have been established that characterize fluctuations of turbulent velocity within the so-called inertial range of scales. Subsequently, Obukhov \cite{Obuk49} and Corrsin \cite{Corr51} expanded this approach to include fluctuations of a passive scalar, such as temperature or impurity density, in a turbulent medium. These findings can also be applied to describing the statistical properties of fluctuations of the refractive index.

The problems related to the light propagation in a turbulent medium were intensively studied in the second half of the twentieth century. The results of the investigations are outlined in Refs. \cite{Tat67,Goodman,Tat75,Tat751,StrohbehnBook1978,AP98}. Recently, interest in the problem has been renewed, mainly due to the increasing capabilities of numerical simulations, which allow one to obtain detailed information about light propagation in a turbulent medium, see Refs. \cite{Vorontsov2010,Vorontsov2011,Lachinova2016,Lushnikov2018,Aks19}.

One of the remarkable properties of light beams propagating through turbulent media is the anomalously high probability of events characterized by high intensity $I$, much larger than its mean value. Quantitatively, the events can be described by ``fat'' tails of the probability density function of $I$, characterized by stretched exponents \cite{KL25Uspekhi}. The role of the events increases as the path traveled by the beam grows.

During optical observations, distortions caused by atmospheric turbulence can be partially corrected using a technique called adaptive optics, see, e.g., Refs. \cite{Beck93,Tys10}. In Ref. \cite{BKLS24}, we have proposed a new method for measuring the Fried parameter, $r_0$, see Ref. \cite{Fried75}. The parameter is a crucial characteristic of light in the turbulent atmosphere determining the requirements for spatial resolution in adaptive optics systems. The idea of Ref. \cite{BKLS24} is in exploiting the off-diagonal component of the phase gradient correlation function, which can be measured, say, using a Shack-Hartmann wavefront sensor \cite{Shack01}.

The expressions presented in Ref. \cite{BKLS24} were derived for the case where the separation $r$ between the observation points is larger than the radius of the first Fresnel zone. The results reported in Ref. \cite{KLS25} expand the approach to arbitrary separations $r$. Corrections to the zeroth approximation, used in Ref. \cite{KLS25} are discussed in Ref. \cite{KL26a}. The results of this work show that the perturbation series for correlation functions of the phase gradients is uniform and is controlled by the dimensionless Rytov variance $\sigma_R^2$, unlike the perturbation series for the envelope $\Psi$, which is non-uniform.

As it was argued in Ref. \cite{Tat67}, at large values of $\sigma_R^2$, the envelope of the beam follows Gaussian statistics. The pair correlation function of the envelope was found in Refs. \cite{Shis68,Tat69}. We demonstrate that the correlation function of the phase gradients in the Gaussian case can be expressed in terms of the correlation function of the envelope.

To verify the analytical predictions, we conducted direct numerical simulations of light propagation in turbulent media, see Ref. \cite{BKLNS26a}. These simulations were performed using the traditional approach \cite{AP98}, where a continuous turbulent medium is modeled using a series of screens, on which random phase shifts of the light wave occur. The statistics of these phase shifts is chosen to mimic the effect of continuous fluctuations in the refractive index on the light wave. The results reported in Ref. \cite{BKLNS26a} are in a good agreement with the theoretical predictions.

We extended our simulations to the case of strong scintillations, aiming to measure the pair correlation function of the phase gradients. The results of the simulations cannot be explained in the framework of a perturbation series. From the other side, we know that the envelope of the wave packet has Gaussian statistics in the regime of strong scintillations \cite{Tat67}. The property allows one to express the pair correlation function of the phase gradients in terms of the pair correlation function of the envelope, thus obtaining an explicit expression for the former. It is in good agreement with the results of our direct numerical simulations.

The structure of our paper is as follows. Section \ref{sec:mainrel} contains the basic relations concerning the light propagation in a turbulent medium, including fluctuations of the refractive index, the envelope of the wave packet. Section \ref{sec:gausspsi} contains a derivation of the expression for the pair correlation function of the phase gradients in terms of the pair correlation function of the envelope. Section \ref{sec:numerics} is devoted to describing results of our numerical simulations conducted for the initial plane wave. Here we compare our numerical results to analytical ones. In Section \ref{sec:conclusion}, we summarize our findings and present the main conclusions of our research.

\section{Main relations}
\label{sec:mainrel}

We consider light wave packets propagating in a turbulent medium, such as the atmosphere, assuming that we are dealing with a monochromatic wave. These wave packets can be characterized by their envelope, $\Psi$, which describes the structure of the wave packet on scales larger than its wavelength. The envelope $\Psi$ is a complex field determining, say, the electric field, which is proportional to
\begin{equation}
\mathrm{Re}\, [\Psi \exp(ik_0z)],
\nonumber
\end{equation}
where $k_0$ is the wave vector of the carrying wave and $z$ is the coordinate in the direction of the wave packet propagation.

The envelope $\Psi$ is controlled by the following parabolic equation
\begin{equation}
i \partial_z \Psi +\frac{1}{2k_0}\nabla^2 \Psi
+k_0 \nu  \Psi=0.
\label{gain1}
\end{equation}
Here $\nu$ designates a fluctuation of the refraction index and the vector operator $\nabla$ in Eq. (\ref{gain1}) designates the coordinate gradient in the direction perpendicular to the $Z$-axis, $\nabla=(\partial_x,\partial_y)$. The equation (\ref{gain1}) implies that the envelope, $\Psi$, adapts simultaneously to the state of the medium, which is described by the field $\nu$. This property is justified by the speed of light being very high.

A theoretical investigation of light propagation based on the equation (\ref{gain1}) involves, generally, several steps. First of all, one introduces a source generating some initial envelope $\Psi_0(x,y,0)$ at $z=0$. Then one solves the evolution equation (\ref{gain1}) along $z$ with the initial condition $\Psi_0(x,y,0)$ at a given field $\nu$. And, finally, one finds correlation functions of $\Psi,\Psi^\star$ (where star designates complex conjugation) at the final $z$, where a receiver is placed. The correlation functions, found by averaging over realizations of $\nu$, contain the complete information about statistical properties of the light beam at the receiver.

Of course, it is impossible to conduct the program explicitly. That is why some approximate methods are used to evaluate the correlation functions. One such method is the perturbation theory used in the regime of weak scintillations. In the following, we discuss the conditions under which the perturbation theory can be applied. Let us first introduce some quantities that will be needed for this discussion.

Statistical properties of the refraction index $\nu$ can be characterized by its second order structure function. In the inertial range of turbulence the structure function follows a power law
\begin{eqnarray}
\langle [\nu  (\bm r_1,z_1) -\nu  (\bm r_2,z_2)]^2\rangle
=C_n^2 (r^2+z^2)^{1/3},
\label{KolmOb}
\end{eqnarray}
characteristic of a passive scalar, see Refs. \cite{Obuk49,Corr51}. Here and below the angular brackets denote time averaging. In the expression (\ref{KolmOb}), the quantity $C_n^2$ determines the strength of the fluctuations of the refractive index, $\bm r=\bm r_1-\bm r_2$, $z=z_1-z_2$. The expression (\ref{KolmOb}) suggests that the turbulent fluctuations in the inertial range are statistically homogeneous and isotropic, which is consistent with experimental observations \cite{Monin,Frisch}.

We assume that the propagation distance of the wave packet is much larger than all its characteristic lateral scales. Then, the field $\nu$ can be treated as a quantity which is short correlated along the $Z$-axis. The property motivates the approximation
\begin{eqnarray}
\langle \nu  (\bm r_1,z) \nu  (\bm r_2,\bm z_1)\rangle=
\delta(z-z_1) C_n^2 {\mathcal A}(\bm r),
\label{delta}
\end{eqnarray}
where $\bm r=\bm r_1-\bm r_2$. The factor $C_n^2$ in Eq. (\ref{delta}) can be thought as a gradually varying function of $z_1$. The characteristic scale of the variations should be larger than the lateral scales.

The function $\mathcal A$ in Eq. (\ref{delta}) is determined by the spectrum of the fluctuations of the refractive index. The function ${\mathcal A}(r)$ is often chosen to be formed by the von Karman spectrum \cite{Kar48}
\begin{equation}
{\mathcal A}(r)=
\int \frac{d^2 q}{(2\pi)^2}
\exp(i\bm q \bm r) \frac{8.19}{[q^2+(2\pi/L_0)^2]^{11/3}},
\label{screen2}
\end{equation}
where $L_0$ is the outer scale of turbulence. If $r\ll L_0$ then \cite{BKLS24}
\begin{equation}
{\mathcal A}(\bm r)=
{\mathcal A}_0  -1.4572\, r^{5/3}
+ 1.1727 \kappa^{1/3} r^2,
\label{corrfr2}
\end{equation}
where ${\mathcal A}_0\sim L_0^{5/3}$. The expression (\ref{corrfr2}) represents the leading order terms in the small parameter $r/L_0$.

The correlation length of the envelope $\Psi$ in the transverse direction, the Fried parameter $r_0$, is defined in accordance with Ref. \cite{Fried75}
\begin{equation}
r_0^{-5/3}= 0.423 \, k_0^2 \int_0^z d\zeta\, C_n^2(\zeta).
\label{Fried}
\end{equation}
For the homogeneous medium where $C_n^2$ is independent of $z$
\begin{equation}
r_0^{-5/3}= 0.423 \, k_0^2 z C_n^2.
\label{Fried2}
\end{equation}
The value of $r_0$ diminishes as the distance $z$ travelled by the wave packet grows.

Traditionally, the level of scintillations is characterized by a dimensionless parameter called the Rytov variance $\sigma_R^2$. For a perturbed plane wave the Rytov variance is defined as
\begin{equation}
\sigma_R^2=2.25 \, k_0^{7/6} z^{5/6}
\int_0^z d\zeta\, (1-\zeta/z)^{5/6} C_n^2(\zeta),
\label{pla2}
\end{equation}
see, e.g., Ref. \cite{Spencer2021}. For the homogeneous medium one finds
\begin{equation}
\sigma_R^2=1.23\,  C_n^2 k_0^{7/6} z^{11/6},
\label{rytovd}
\end{equation}
after taking the integral in Eq. (\ref{pla2}). The quantity $\sigma_R^2$ grows as the wave packet runs.

One of the quantities that characterizes the fluctuations of the envelope, $\Psi$, is its pair correlation function
\begin{equation}
F(\bm r_1,\bm r_2,z)=
\langle \Psi(\bm r_1,z)
\Psi^\star(\bm r_2,z) \rangle.
\label{paircorrfu}
\end{equation}
The equation for the pair correlation function (\ref{paircorrfu}) was obtained and analysed in Refs. \cite{Shis68,Tat69}. Let us emphasize that the analysis is not sensitive to the value of $\sigma_R^2$ and provides the pair correlation function for both the regimes of weak and strong scintillations.

For the initial plane wave passing the turbulent path, where fluctuations of the refractive index are determined by the von Karman spectrum, the normalized pair correlation function is determined by the expression
\begin{eqnarray}
\ln \left[F(r)/F(0)\right] =
- 3.44  \left(\frac{r}{r_0}\right)^{5/3}
\nonumber \\ \times
\left[1-1.485 \left(\frac{r}{L_0}\right)^{1/3}\right].
\label{BKLS24}
\end{eqnarray}
The pair correlation function only depends on the distance between observation points $r$, $\bm r=\bm r_1-\bm r_2$. The expression (\ref{BKLS24}) is correct at small $r/L_0$, see Ref. \cite{BKLS24}.

With applications to adaptive optics in mind, we focus on analyzing the pair correlation function of phase gradients
\begin{equation}
Q_{\alpha\beta}=\langle \partial_\alpha \varphi (\bm r_1,z)
\partial_\beta \varphi(\bm r_2,z) \rangle,
\label{kgrph1}
\end{equation}
where $\varphi$ is the phase of the envelope $\Psi$, $\Psi=|\Psi|\exp(i\varphi)$. Just the quantity (\ref{kgrph1}) is measured by a Shack-Hartmann wavefront sensor \cite{Shack01}.

If $\sigma_R^2$ is large, then the envelope $\Psi$ can be thought of as a sum of many terms, resulting from diffraction on random fluctuations in the refractive index. Therefore, the envelope $\Psi$ has Gaussian statistics due to the central limit theorem, see, e.g., Ref \cite{Tat67}. Then statistical properties of $\Psi$ are completely determined by the pair correlation function (\ref{paircorrfu}). The applicability conditions for the Gaussian statistics of $\Psi$ are discussed in the works \cite{ZKT77,Char94,KLL20}. It is valid for correlation functions of $\Psi$ of order less than some number $n_c$ which grows as $z$ increases.

\section{Analytic derivation}
\label{sec:gausspsi}

Here we analytically examine the case of large values of $\sigma_R^2$ where the envelope $\Psi$ has Gaussian statistics in the main approximation. Then, the only factor that determines the statistical properties of $\Psi$ at a given $z$ is its pair correlation function (\ref{paircorrfu}). Further we express the pair correlation function of the phase gradients via the pair correlation function (\ref{paircorrfu}).

The starting point of our calculations is the following expression for the pair correlation function of the phase gradients (\ref{kgrph1})
\begin{eqnarray}
Q_{\alpha\beta}=\frac{1}{4}\left\langle
\frac{J_\alpha(\bm r_1) J_\beta(\bm r_2)}{|\Psi(\bm r_1)|^2 |\Psi(\bm r_2)|^2}
\right\rangle,
\label{kgrph2}
\end{eqnarray}
where
\begin{eqnarray}
J_\alpha=-i(\Psi^\star\partial_\alpha \Psi -\partial_\alpha \Psi^\star \Psi).
\label{kgrph3}
\end{eqnarray}
The average (\ref{kgrph2}) can be rewritten as the following integral
\begin{eqnarray}
Q_{\alpha\beta}=\frac{1}{4}
\iint_0^\infty dt\, ds\, \left\langle {\mathcal E}
J_\alpha(\bm r_1) J_\beta(\bm r_2) \right\rangle,
\label{kgrph4} \\
{\mathcal E}=
\exp\left[-t |\Psi(\bm r_1)|^2 -s |\Psi(\bm r_2)|^2\right]  ,
\label{kexponent}
\end{eqnarray}
which is more convenient for subsequent handling.

Further, having in mind the case of the initial plane wave, we assume that the pair correlation function $F$ (\ref{paircorrfu}) only depends on the difference $\bm r=\bm r_1-\bm r_2$. The property reflects homogeneity of the problem in the transverse planes. Moreover, $F$ is real in this case, see Eq. (\ref{BKLS24}). We accept the normalization $F(0)=1$. However, we would like to stress that our procedure can be easily generalized for arbitrary $F$.

To calculate the average of products of a field with Gaussian statistics, Wick's theorem can be used \cite{Wick}. The theorem states that if a field $f(\bm r)$ has Gaussian statistics then the average of a product $f(\bm r_1) f(\bm r_2) \dots$ is equal to the sum of terms in which the product $f(\bm r_1) f(\bm r_2) \dots$ is substituted by the product of the pair averages like $\langle f(\bm r_1) f(\bm r_2)\rangle \langle f(\bm r_3) f(\bm r_4)\rangle \dots$. One should take into account all possible variants of pairing. Thus, Wick's theorem is purely combinatoric.

For the subsequent analysis, we introduce averages with the additional factor (\ref{kexponent}) incorporated into the probability density function of $\Psi$. We denote such averages using the floor brackets:
\begin{equation}
\langle {\mathcal E} B \rangle
={\mathcal Z} \lfloor B \rfloor, \quad
{\mathcal Z} =\langle {\mathcal E} \rangle.
\label{kfloors}
\end{equation}
Here, $B$ is a function of $\Psi$, $\Psi^\star$, and their gradients, and ${\mathcal Z}$ is the correction to the normalization factor. Introducing the additional factor (\ref{kexponent}) into the probability density still results in a Gaussian probability distribution for $\Psi$. Therefore the new averages (designated by floors) are subject to Wick's theorem as well.

To relate the ``new'' and ``old'' averages, $\lfloor B \rfloor$ and $\langle B \rangle$, one should expand the exponential ${\mathcal E}$ in the expression $\langle{\mathcal E B}\rangle$ and then use Wick's theorem. Applying the procedure to the averages $\langle {\mathcal E} \Psi_1 \Psi_2^\star \rangle$, $\langle {\mathcal E} |\Psi_2|^2 \rangle$, we find that $\lfloor \Psi_1 \Psi_2^\star \rfloor$, $\lfloor |\Psi_2|^2 \rfloor$ are reduced to geometrical progresses, all terms of which are expressed in terms of the old averages. Here and below we use for brevity the designations like $\Psi_1=\Psi(\bm r_1)$, $\Psi_2=\Psi(\bm r_2)$.

Summation of the geometrical progresses leads to the closed system of equations
\begin{eqnarray}
\lfloor |\Psi_2|^2 \rfloor =
1-t F \lfloor \Psi_1 \Psi_2^\star \rfloor
-s \lfloor |\Psi_2|^2 \rfloor,
\label{kgrph6} \\
\lfloor \Psi_1 \Psi_2^\star \rfloor
=F -t \lfloor \Psi_1 \Psi_2^\star \rfloor
-s F \lfloor |\Psi_2|^2 \rfloor.
\label{kgrph7}
\end{eqnarray}
The equations (\ref{kgrph6},\ref{kgrph7}) can easily be solved to obtain
\begin{eqnarray}
\lfloor |\Psi_2|^2 \rfloor =
\frac{1+t(1-F^2)}{1+t+s +ts(1-F^2)},
\label{kgrph8} \\
\lfloor \Psi_1 \Psi_2^\star \rfloor
=\frac{F}{1+t+s +ts(1-F^2)}.
\label{kgrph9}
\end{eqnarray}
Analogously one finds the expression
\begin{equation}
\lfloor |\Psi_1|^2 \rfloor =
\frac{1+s(1-F^2)}{1+t+s +ts(1-F^2)},
\label{kgrph10}
\end{equation}
that can be obtained also from Eq. (\ref{kgrph8}) by permuting $t \leftrightarrow s$.

Taking the derivatives of
\begin{equation}
{\mathcal Z}(t,s)=\left\langle
\exp\left[-t |\Psi(\bm r_1)|^2 -s |\Psi(\bm r_2)|^2\right] \right\rangle,
\nonumber
\end{equation}
one finds
\begin{equation}
\frac{\partial\ln{\mathcal Z}}{\partial t}
=-\lfloor |\Psi_1|^2 \rfloor, \quad
\frac{\partial\ln{\mathcal Z}}{\partial s}
=-\lfloor |\Psi_2|^2 \rfloor.
\nonumber
\end{equation}
Substituting here the expressions (\ref{kgrph8},\ref{kgrph10}) and solving the resulting differential equations one finds the explicit expression
\begin{equation}
{\mathcal Z}=
\frac{1}{1+t+s +ts(1-F^2)}.
\label{kgrph13}
\end{equation}
The constant of integration here is fixed by the condition ${\mathcal Z}(0,0)=1$.

One can include the derivatives of the envelope $\Psi$ into the developed scheme. By applying the same procedure, we find the relations
\begin{eqnarray}
\lfloor \partial_\alpha \Psi_1 \Psi_2^\star \rfloor
=\partial_\alpha F -s \partial_\alpha F \lfloor |\Psi_2|^2\rfloor
\nonumber \\
=\frac{1+t}{1+t+s +ts(1-F^2)} \partial_\alpha F,
\label{kgrph14} \\
\lfloor  \Psi_1 \partial_\alpha\Psi_2^\star \rfloor
=-\partial_\alpha F + t \lfloor |\Psi_1|^2\rfloor\partial_\alpha F
\nonumber \\
=-\frac{1+s}{1+t+s +ts(1-F^2)} \partial_\alpha F,
\label{kgrph15}
\end{eqnarray}
analogously to deriving Eqs. (\ref{kgrph8},\ref{kgrph9},\ref{kgrph10}). Next, one finds
\begin{eqnarray}
\lfloor  \partial_\alpha\Psi_1 \Psi_1^\star \rfloor
=\lfloor \Psi_1  \partial_\alpha \Psi_1^\star \rfloor
= -s \lfloor \Psi_1^\star \Psi_2 \rfloor \partial_\alpha F
\nonumber \\
=-\frac{s F}{1+t+s +ts(1-F^2)}\partial_\alpha F.
\label{kgrph16}
\end{eqnarray}
The last object needed for our calculations is
\begin{eqnarray}
\lfloor \partial_\alpha \Psi_1 \partial_\beta \Psi_2^\star \rfloor
=-\partial_\alpha\partial_\beta F
+t\lfloor  \partial_\alpha \Psi_1 \Psi_1^\star \rfloor
\partial_\beta F
\nonumber \\
=-\partial_\alpha\partial_\beta F
-\frac{t s F \partial_\alpha F \partial_\beta F}{1+t+s +ts(1-F^2)}.
\label{kgrph17}
\end{eqnarray}
Thus we express all new pair correlation functions via the old ones.

Now we return to the expression (\ref{kgrph4}). The average can be written as
\begin{eqnarray}
Q_{\alpha\beta}=
\frac{1}{2}\iint_0^\infty dt\, ds\,
{\mathcal Z} \qquad
\nonumber \\
\left(\lfloor \Psi_1 \Psi_2^\star \rfloor
\lfloor \partial_\alpha \Psi_1^\star \partial_\beta \Psi_2 \rfloor
-\lfloor \Psi_1^\star \partial_\beta \Psi_2 \rfloor
\lfloor \partial_\alpha \Psi_1 \Psi_2^\star \rfloor\right),
\label{kwickth}
\end{eqnarray}
where we used Wick's theorem for the new averages. Note that $\lfloor J_\alpha \rfloor=0$ thanks to Eq. (\ref{kgrph16}). Substituting the expressions (\ref{kgrph9},\ref{kgrph13},\ref{kgrph14},\ref{kgrph15},\ref{kgrph17}) into Eq. (\ref{kwickth}), one obtains
\begin{eqnarray}
Q_{\alpha\beta}
=\frac{1}{2} \iint_0^\infty dt\, ds\, {\mathcal Z}^2
(\partial_\alpha F \partial_\beta F -F \partial_\alpha \partial_\beta F)
\nonumber \\
=\frac{1}{2}
\partial_\alpha \partial_\beta \ln F
 \ln(1-F^2).
\label{kgrph18}
\end{eqnarray}
Thus, we expressed the pair correlation function of the phase gradients (\ref{kgrph1}) via the pair correlation function of the envelope (\ref{paircorrfu}).

\section{Numerics}
\label{sec:numerics}

In this section we present results of our numerical simulations. Our setup corresponds to a monochromatic light wave with a wavelength of $\lambda = 0.55 \mu m$, which propagates through a statistically homogeneous medium. We use computational box which is a square $1.28\times 1.28$ meters. The initial profile of the envelope, $\Psi$, had a flat central region with a diameter of $1$ meter and super-Gaussian wings. The width of the box is much greater than the size of the inhomogeneities associated with fluctuations of the refractive index. Thus, in the region near the axis of the beam the regime of a plane wave is realized.

The beam propagation is assumed to be governed by the parabolic equation (\ref{gain1}). Numerically, the equation (\ref{gain1}) is solved by the finite difference method with splitting diffraction and refraction processes, see, e.g., Ref. \cite{Fle76,Kan96}. The continuous turbulent medium is modeled by a chain of infinitely thin phase screens that are perpendicular to the direction of beam propagation. The computation process involves a series of consecutive transformations of the envelope $\Psi$, including the addition of the phase jumps at the screens and the free diffraction of $\Psi$ between them.

The diffraction of the beam in the regions between the screens is computed by solving the propagation equation with the Ladagin \cite{Lad85} difference scheme, which has zero amplitude error and a fourth-order phase error when integrating the diffraction operator. A number of the screens was chosen in such a way that, at the evolution between the screens, the phase shift due to diffraction was small, and the phase dispersion on the screen did not exceed $1$ rad$^2$. The number varied in the range $100 \div 200$, depending on the length of the path. The grid pitch was taken from the range $0.05 \div 0.25$ cm, its value was selected to successfully solve the equation and was dependent on the distance traveled by the beam.

The envelope is transformed on a screen according to the rule
\begin{equation}
\Psi_+(x,y)=\exp[i\Delta\varphi(x,y)]\Psi_-(x,y).
\label{transf}
\end{equation}
Here $\Psi_-$ is the envelope before the screen, $\Psi_+$ is its value after the screen, and $\Delta\varphi$ is the phase jump. The phase jumps for different screens are chosen to be independent random functions. For each screen it is generated to ensure the pair correlation function of the phase jump
\begin{equation}
\langle \Delta\varphi(\bm r_1) \Delta\varphi(\bm r_2) \rangle
=   C_n^2 k_0^2 {\mathcal A}(\bm r) \Delta z,
\label{screen}
\end{equation}
where $\bm r=\bm r_1-\bm r_2$ and $\Delta z$ is the separation between the screens. The expression (\ref{screen}) represents the integral influence of the random diffraction index $\nu$ in the layer of width $\Delta z$ in accordance with Eq. (\ref{delta}).

The function ${\mathcal A}(r)$ was chosen to be formed by the von Karman spectrum (\ref{screen2}) with $L_0=20$ m, where the integration over $\bm q$ was substituted by a sum over harmonics. The generation of random phase jumps $\Delta\varphi$ on the screens was carried out using a combination of the usual spectral method and the method of subharmonics \cite{Kan98,Joh94}. The idea of the method is to condense the nodes of the computational grid in the spectral plane near the origin, by adding extra harmonics (or subharmonics) to the standard set.

The addition of the subharmonics to the spatial spectrum is done iteratively. At each iteration, extra $32$ harmonics are added to the phase spectrum with wave vectors three times smaller than the previous one (for details see Ref. \cite{Bog24}). In this way, it is possible to reproduce correctly the large-scale spatial behavior of the pair correlation function (\ref{screen}) without increasing the size of the box. In our computations, the number of iterations ranged from $5$ to $7$.

\begin{figure}
\begin{center}
\includegraphics[width=0.5\textwidth]{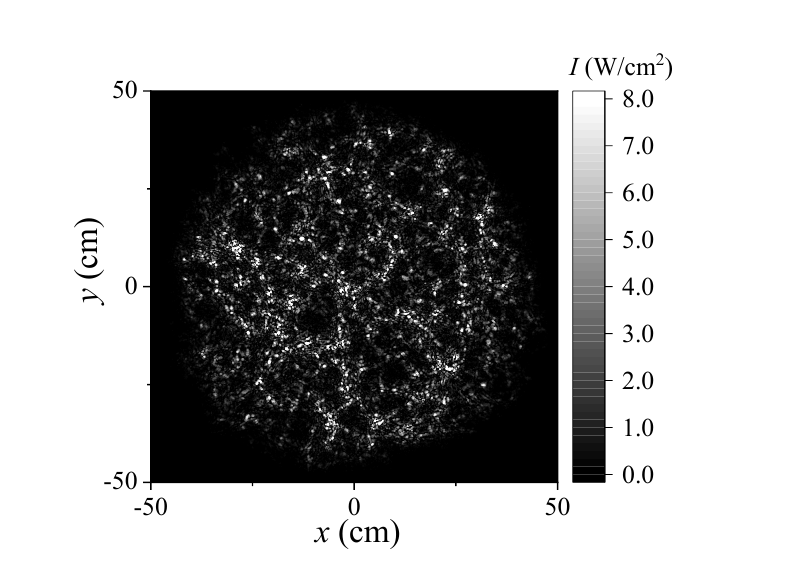}
\includegraphics[width=0.5\textwidth]{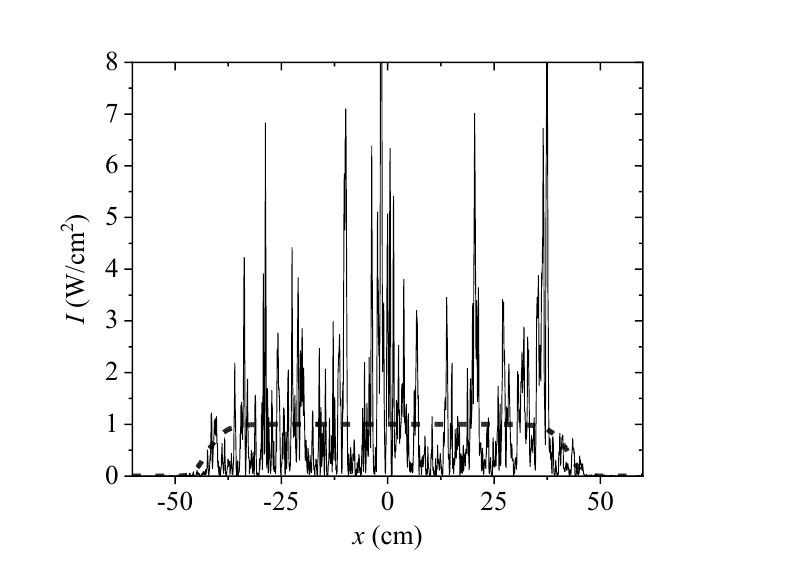}
\end{center}
\caption{The output intensity as a function of $x,y$ (the upper graph) and its dependence on $x$ at $y=0$ (the lower graph) for $\sigma_R^2 =5.73$, $r_0=0.62$ cm.
The dashed line represents the ideal intensity profile that would be observed in the absence of turbulence.}
\label{fig:2-1}
\end{figure}

At the end of the turbulent route, containing by the random screens, a profile of the envelope $\Psi$ was recorded. An example of the output intensity $I=\Psi^\star \Psi$, as a function of a coordinate, is depicted in Fig. \ref{fig:2-1}. During the repeated passage of the wave along the implementations of the turbulent route by the Monte Carlo method, a set of $\Psi$ was formed. In our simulations, the size of the set ranged from $300$ to $500$. Using the set, one can find some averaged characteristics of the light beam.

\begin{figure}
\begin{center}
\includegraphics[width=0.5\textwidth]{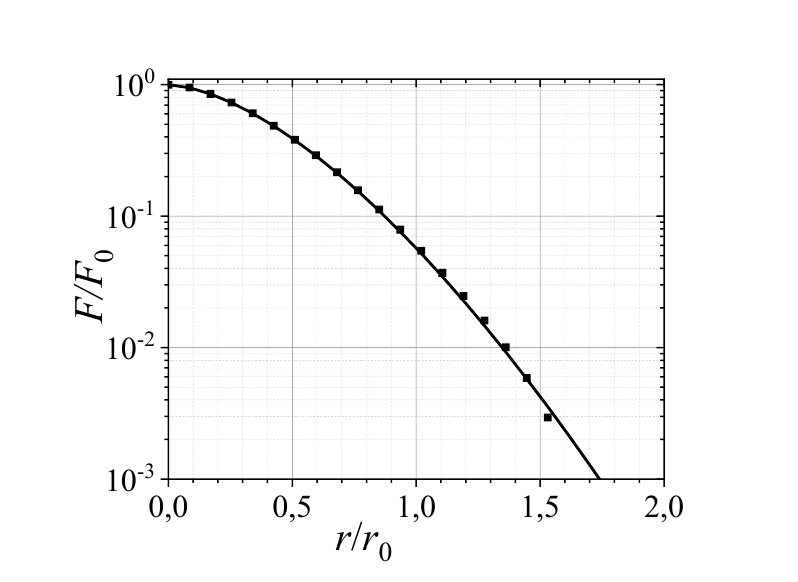}
\includegraphics[width=0.5\textwidth]{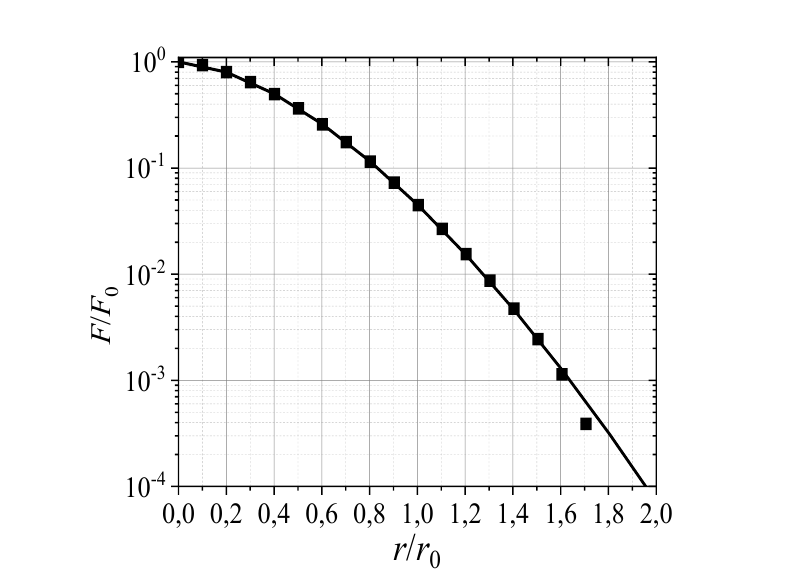}
\end{center}
\caption{The normalized pair correlation function of the envelope (\ref{paircorrfu}) as a function of $r/r_0$ for $\sigma_R^2=1.05$, $r_0=2.94$ cm (the upper graph) and $\sigma_R^2=5.73$, $r_0=0.62$ cm (the lower graph. The solid line is drawn using the analytic expression (\ref{BKLS24}) and the filled squares represent the numerical data extracted from the simulations.}
\label{fig:2-2}
\end{figure}

The accuracy of the computations was controlled as follows. For a set of transverse distributions $\Psi(x.y)$, recorded at the end of the path, the pair correlation function (\ref{paircorrfu}) was calculated by averaging over the set. For statistically homogeneous and isotropic turbulence, the pair correlation function depends only on the distance between the points, $\bm r=\bm r_1-\bm r_2$, and is real. The properties are revealed after averaging, indeed. It confirms validity of our computational scheme.

\begin{figure}
\begin{center}
\includegraphics[width=0.5\textwidth]{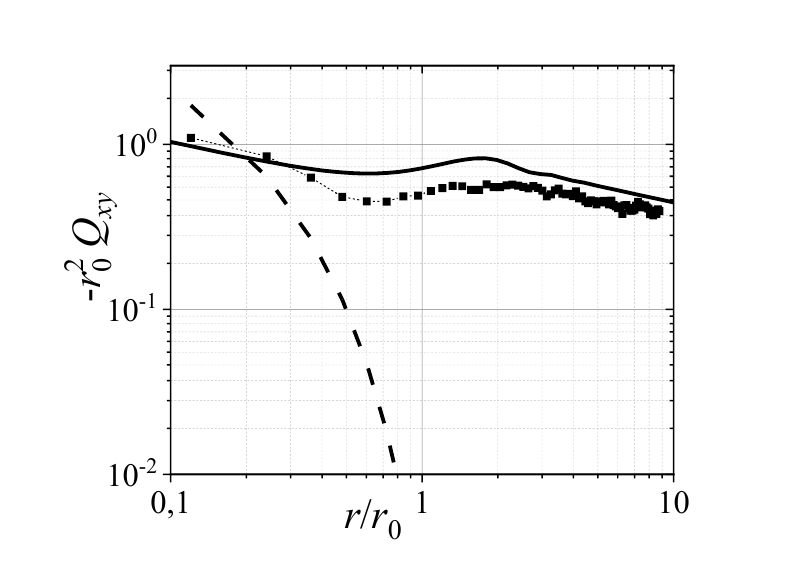}
\includegraphics[width=0.5\textwidth]{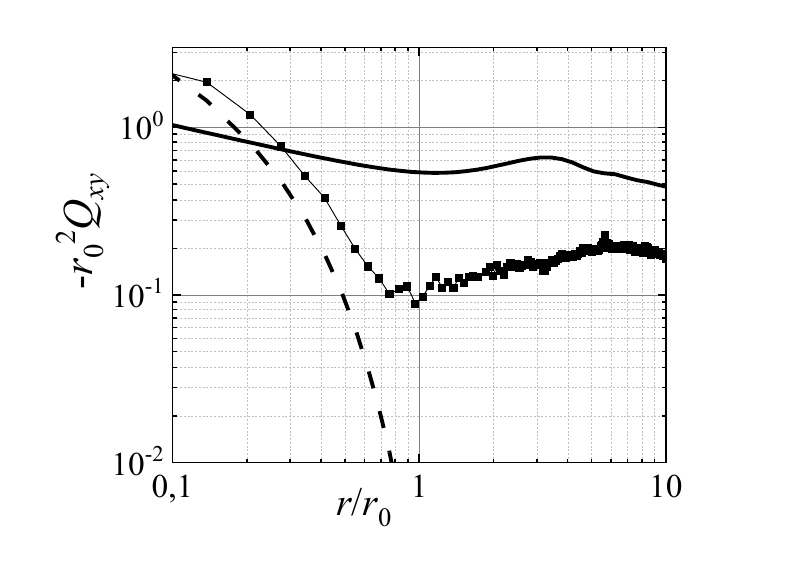}
\includegraphics[width=0.5\textwidth]{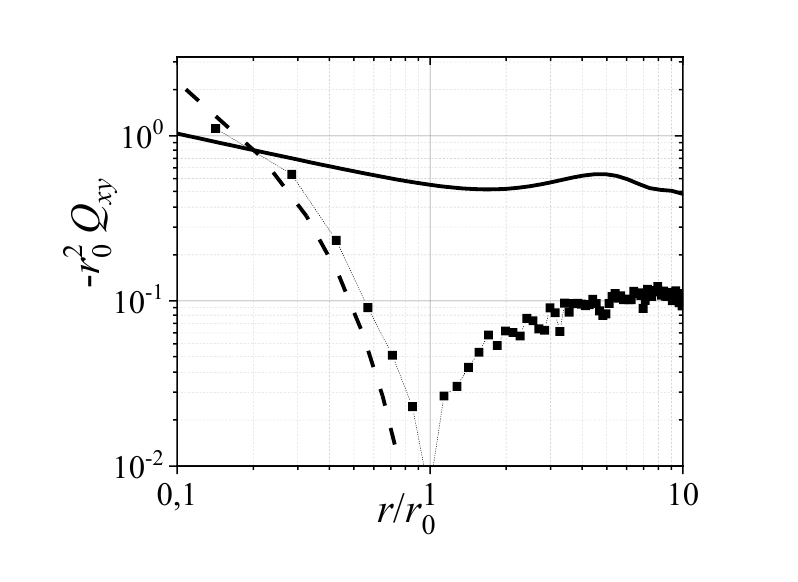}
\end{center}
\caption{The off-diagonal component $Q_{xy}$ of the pair correlation of the phase gradients (\ref{kgrph1}) multiplied by $-r_0^2$ taken at $x=y$ as a function of $r/r_0$, for
$\sigma_R^2=1.05$, $r_0=2.94$ cm (the upper graph),
$\sigma_R^2=3.1$, $r_0=1.28$ cm (the middle graph),
$\sigma_R^2=5.73$, $r_0=0.62$ cm (the lower graph).
The solid line represents the analytical expression obtained from the zeroth order of the perturbation theory, the dashed line is drawn by using the expression (\ref{kgrph18}), and the filled squares represent the numerical data extracted from the simulations.}
\label{fig:2-3}
\end{figure}

A comparison of the analytical expression (\ref{BKLS24}) and the numerical value extracted from simulations for specific values of the parameters $\sigma_R^2$, $r_0$ and $L_0=20$ m is shown in Fig. \ref{fig:2-2}. It can be seen that the numerical data reproduce the expression (\ref{BKLS24}) pretty well, which indicates that all the main turbulence parameters are adequately reproduced in our simulations.

Now we pass to the pair correlation function of the phase gradients $Q_{\alpha\beta}$. The analytical expression for the pair correlation function in zeroth order of the perturbation theory is given in our paper \cite{KLS25}. The off-diagonal component $Q_{xy}$ calculated in this approximation is drawn by the solid line. The dashed line represents the expression (\ref{kgrph18}), derived for the Gaussian statistics of the envelope $\Psi$.

We see that as $\sigma_R^2$ increases, the profile of $Q_{xy}$ gradually evolves from the zeroth approximation to the value determined by Eq. (\ref{kgrph18}), as expected. However, it is true only for $r\lesssim r_0$. For $r\gtrsim r_0$ the behavior of $Q_{xy}$ is quite different, it grows as $r$ increases approaching the zeroth order value. The fact requires a special explanation, which will be provided in the following subsection.

\subsection{Explanation}
\label{subsec:explanation}

Gaussianity of $\Psi$ implies that the main contribution, say, to the fourth correlation function $\langle \Psi_1\Psi_2\Psi_3^{\star}\Psi_4^{\star} \rangle$ is given by the products of the pair correlation functions defined by Eq. (\ref{paircorrfu}). Corrections determined by the irreducible part of the fourth correlation function are small when the distances between the points are $r \lesssim r_0$, provided $\sigma^2_R$ is large, see Refs. \cite{ZKT77,Char94,KLL20}.

However, the pair correlation function (\ref{paircorrfu}) decays rapidly as the distance $r$ between the points increases in the region where $r \gtrsim r_0$, see Eq. (\ref{BKLS24}). In contrast to the behavior, the irreducible fourth correlation function decreases only as a power of $r$ with increasing $r$, as demonstrated in Ref. \cite{Masnev26}. Therefore, the irreducible contribution quickly becomes larger than the reducible contribution as $r$ increases. This implies that the statistics of $\Psi$ at $r \gtrsim r_0$ cannot be treated as Gaussian.

The graphs in Fig. \ref{fig:2-3} show that the off-diagonal part of the pair correlation function of phase gradients approaches the zeroth order approximation for large distances between the observation points. The peculiarity is in accordance with the calculations presented in Ref. \cite{KL26a} where the regular perturbation theory in terms of $\ln \Psi$ has been developed. It was demonstrated in the work that perturbation corrections to the zeroth order approximation remain small at large separations even at large values of the Rytov dispersion $\sigma_R^2$.

Thus, the $r$-dependence of $Q_{xy}$ obtained numerically is in accordance with the theoretical concepts.

\section{Conclusion}
\label{sec:conclusion}

We theoretically and numerically examined the light beam propagating through a turbulent medium in the regime of strong scintillations. The medium is characterized by Kolmogorov's statistics of the refractive index fluctuations, homogeneous in space. Realizations of the refractive index are modeled by a large number of screens, ranging from $100$ to $200$, where the phase of the beam's envelope $\Psi$ experiences random jumps. The propagation of the beam between the screens is free.

The simulations allowed us to extract the pair correlation function of the phase gradients at the end of the beam path. As expected, the results show significant deviations from the zeroth order of perturbation theory, which increase as the Rytov variance, $\sigma^2_R$, grows. We compared the results with the analytical expression for the pair correlation function of the phase gradients, which was derived under the assumption of Gaussian statistics for the envelope $\Psi$, see Section \ref{sec:gausspsi}. The comparison shows a good agreement between the theory and the numerical results at the highest value of $\sigma_R^2$ and distances $r \lesssim r_0$. However, for larger $r$ the agreement is violated. The explanation of this fact is given in subsection \ref{subsec:explanation}.

Looking on the graphs in Fig. \ref{fig:2-3} we conclude that in the strong scintillations regime the off-diagonal part of the pair correlation function of phase gradients has a minimum near the Fried parameter $r_0$. The property enables one to evaluate the Fried parameter using observable data in the regime of strong scintillations.

Further, we plan to conduct experimental observations of the phase correlations to confirm our analytical and numerical findings. It should be noted that investigating the behavior of phase correlations is challenging experimentally in the strong scintillation regime.

It is well known that under certain conditions, atmospheric turbulence can be non-Kolmogorov, see, e.g., Ref. \cite{Korot21}. The property is usually related to the breaking of the isotropy of turbulence, which is a complex phenomenon that requires a special investigation that lies outside the scope of this work.

\acknowledgements

This work was supported by the scientific program of the National Center for Physics and Mathematics of RF, Section 4, stage 2026-2028.

\end{document}